Elham Heidari, Hamed Dalir*, Moustafa Ahmed, Volker J. Sorger, Ray T. Chen

# Hexagonal Transverse Coupled Cavity VCSEL Redefining the High-Speed Lasers

**\*Corresponding author: Hamed Dalir**
Omega Optics, Inc. 8500 Shoal Creek Blvd., Bldg. 4, Suite 200, Austin, Texas 78757, USA
Email: hamed.dalir@omegaoptics.com

**Elham Heidari, Ray T. Chen:** Microelectronics Research Center, Electrical and Computer Engineering Department, University of Texas at Austin, Austin, Texas 78758, USA

**Moustafa Ahmed:** Department of Physics, Faculty of Science, King Abdul-Aziz University, 80203 Jeddah 21589, Saudi Arabia

**Volker J. Sorger:** Department of Electrical and Computer Engineering, George Washington University, Washington. DC 20052, USA

**Abstract:** The vertical-cavity surface-emitting lasers (VCSELs) have emerged as a vital approach for realizing energy efficient, high speed optical interconnects in the data center and supercomputers. As of today, VCSEL is the most suitable for mass production in terms of cost-effectiveness and reliability. However, there are still key challenges for higher speed modulation above 40 GHz. Here, a hexagonal transverse coupled cavity VCSEL adiabatically coupled through the center cavity is proposed. A 3-dB roll-off modulation bandwidth of 45 GHz is demonstrated, which is five times greater than a conventional VCSEL fabricated on the same epi-wafer structure. While a parity time (PT) symmetry approaches add 'loss' to 'engineer' the topological state of the laser system, here, a radical paradigm shift with gain introduces symmetry breaking. This idea, then enables a single mode operation with a side-mode suppression-ratio (SMSR) of > 30 decibels and signal-to-noise ratio (SNR) of > 45 decibels. The energy distribution inside the coupled cavity system is also redistributed to provide a coherent gain in a spatially separated system. Consequently, throughput power is three times higher than that of the conventional VCSEL.

1. Introduction

Semiconductor lasers allowing for indispensable science and a wide range of technologies, have become one of the most important enablers of photonic-based technologies [1]. VCSELs, a new class of semiconductor lasers [2], specifically, are gaining in importance given their formfactor and opto-electronic performance for their use as an optical source in high speed and short wavelength communications [3,4], and sensors [4-6]. For instance, in recent years VCSELs have been deployed aside semiconductor diode lasers as the sources in cost-effective fiber links and data center networks [7-9] due to their distinct features, such as high reliability, low cost and high yield, low power consumption and easy packaging; low threshold and operating currents, high temperature stability and straightforward fabrication of dense arrays [4,10,11].

For data-transmission applications, a high modulation bandwidth is desired. The 3-dB bandwidth of the VCSEL is limited by thermal effects, parasitic resistance, capacitance, and nonlinear gain effects (such as relaxation oscillations) [12,13]. Thus, with an opportune design of the active region and RC-parasitic, which shunts the modulation current outside the active region at high frequencies, a high modulation bandwidth can be achieved. Optical feedback has proven to increase the modulation bandwidth of VCSELs, due to an induced photon-photon resonance (PPR) effect [14-17]. Dalir et al. [18,19] demonstrated modulation bandwidth enhancement of the VCSEL via adding a single transverse-coupled-cavity (TCC) to a primary VCSEL cavity. The underlying principle is to control the slow-light delay in the TCC via an induced slow-light feedback. However, strong PPR effects incur relatively large kinks in the light-versus-current (L-I) characteristics. The kinks indicate that because of the supply current being increased, the laser behavior turns from stable to unstable [18-22]. Interestingly, multiple TCCs (MTCCs) are advantageous to avoid the kinks in the stable region of L-I curve. This is the common issue in standard coupled cavities (twin cavities) [18,23] as well as optical injection [24].

Here, we demonstrate a micro-cavity laser that is adiabatically coupled with MTCC, forming a hexagonal lattice, which allows slow-light feedback from each TCC directly into the center (modulated) cavity. With an aim to achieve high power, single-mode operation, the Vernier effect in hexagonal VCSEL can be utilized in a larger aperture design [25,26]. We demonstrated operating speeds up-to 45 GHz, side-mode suppression (SMSR) of > 30 dB and signal-to-noise ratio (SNR) of > 45 dB. Finally, we optimized the lattice design to extend the stability along larger output power by about three times when compared with conventional VCSEL [27].

## 2. Structure and Concept

The MTCC VCSEL design aims to facilitate an adiabatically optical energy sharing amongst the coupled cavities with respect to the center lasing cavity (**Figure 1**). Even if this feedback strength of each cavity is only moderate (~ 0.1 THz)[28], due to its adiabatic design, the laser cavity parametrically accumulates an increased amount of the slow-light portion of the light gain and makes this available for the to-be-modulated cavity. The spatial separation of functionality between creating selective gain and modulation functionality is key to realizing both high-speed and high power in single mode operation. This design is advantageous by avoiding the optical loss accumulated in the chain of cascaded TCCs [27].

Unlike a conventional VCSEL design, our MTCC-based structure light has an additional horizontal (lateral) component with an angle close to 90° near the cut-off condition of light propagation [29]. Thus, the main fraction of the slow light effect is totally reflected at the far end of each feedback cavity and is coupled back into the modulating cavity with a coupling ratio of η. The period of each round trip in the TCC of width $L_C$ is $\tau = 2n_g L_c / C$ while $n_g = fn$ is the group index with $n$ and $f$ being the average material refractive index and slow-light factor, respectively [18]. If we consider multiple round trips of the slow light in the MTCCs, the threshold gain $G_{thD}$ of the VCSEL is defined by [30]:

$$G_{th} = G_{thD} - \frac{v_g}{W} \ln \prod_{m=1}^{M} |U_m(t - \tau_m)| \qquad (1)$$

which is generalizations of the forms in [30,31] where $U_m(t - \tau_m)$ is the time-delay function describing the slow-light feedback from the $m^{th}$ TCC. It is defined in terms of the time delay field $E(t - \tau)$ as:

$$U_m(t-\tau_m) = |U_m(t-\tau_m)|e^{j\varphi_m} = 1 + \frac{\eta_m}{\eta_m - 1}\sum_p \sqrt{1-\eta_m^{\ p}}\, e^{-2p\alpha_{cm}L_{cm}} e^{-j2p\beta_{cm}L_{cm}} \sqrt{\frac{S(t-p\tau_m)}{S(t)}} e^{j\theta(t-p\tau_m)-j\theta(t)}$$

(2)

where the summation is over the multiple round trips in TCC, $e^{-2p\alpha_{cm}L_{cm}}$, and $e^{-j2p\beta_{cm}L_{cm}}$ are the loss and phase delay of slow-light during each round trip. $\alpha_{cm} = f\alpha_m$ and $\beta_{cm} = 2\pi n/(\lambda f)$ are the lateral optical loss and propagation constant, where $\alpha_m$ is the material loss related to the $m^{th}$ TCC and $\lambda$ is emission wavelength. In this case, $\eta_m$ is the coupling ratio of the slow-light feedback from the $m^{th}$ cavity, while $L_{cm}$ and $\tau_m$ are the corresponding length and round-trip time, respectively.

The rate equations of the MTCC-VCSEL is given for the injected electron density $N(t)$, photon density $S(t)$ contained in the lasing mode and the optical phase $\theta(t) = arg\ [E(t)]$ as:

$$\frac{dN}{dt} = \frac{\eta_i}{e}I - \alpha v_g \frac{(N-N_T)}{1+\varepsilon S}S - \frac{N}{\tau_e} \quad (3)$$

$$\frac{dS}{dt} = \left[\Gamma a v_g \frac{N-N_{th}}{1+\varepsilon S} - \frac{1}{\tau_p} + \frac{v_g}{W}\sum_{m=1}^{M}\ln|U(t-\tau_m)|\right]S + \Gamma R_{sp} \quad (4)$$

$$\frac{d\theta}{dt} = \frac{\alpha}{2}\left[\Gamma a v_g(N-N_{th}) + \frac{v_g}{W}\sum_{m=1}^{M}\varphi_m\right] \quad (5)$$

where $\eta_i$ represents the injection efficiency, which is the fraction of terminal current provides carriers that recombine in the active region, $\alpha$ is the differential gain of the active region whose volume is $V$, $N_T$ defines the electron numbers at the transparency, and $\varepsilon$ is the gain suppression coefficient. $\Gamma$ and $\tau_p = 1/G_{thD}$ represent the confinement factor and photon lifetime in the lasing cavity, respectively. $\tau_e$ is the electron lifetime due to the spontaneous emission rate, $R_{sp}$ is the spontaneous emission rate, and $N_{th}$ is the electron number at the threshold. We solve these rate equations via the 4$^{th}$ order of Runge-Kutta method using sinusoidal current modulation with bias component $I_b$, modulation component $I_m$, and modulation frequency $f_m$. The integration step is set to 0.2 ps. Six TCCs are considered, M=6. These TCCs are assumed identical having the same length of $L_{cm}=L_c$ and $\tau_m = \tau$ of slow light. A typical limitation of the single TCC is the need for a strong slow-light coupling into the VCSEL cavity to extend the modulation bandwidth further into the mm-waveband [30]. We propose the use of MTCCs in such a way to induce direct slow-light feedback from each TCC. That is, we designed a surrounding MTTCs as shown in the scheme of Figure 1, which would enable for bandwidth enhancement with realistic lower values of the coupling ratio $\eta$.

### 3. Experimental Demonstration of the Hexagonal VCSEL

This formulism can predict the MTCCs small signal response based on the modulation frequency (**Figure 2a**); the laser's intensity modulation (IM) including the multiple PPR effects shows a speed in excess of 40 GHz that is relatively robust against changes in the slow-light feedback ($\eta$). The IM response of a conventional VCSEL ($\eta=0$) shows the expected slow (9 GHz) response (Figure 2a) [18,19, 32]. For a moderate pump and TCC feedback ($I_{laser}=2.5 \times I_{th} = 8\ mA$ and $\eta=0.12$) the IM response exhibits an enhanced carrier to photon resonance (CPR) and multiple PPR before its 3-dB roll-off, which occurs at a bandwidth frequency of 42 GHz (Figure 2a). Increasing the coupling between the inner (modulating) cavity and outer (feedback) cavities exhibits an extended 3-dB roll-off of 100 GHz and beyond by further increasing the coupling strength to $\eta =0.45$ (**Figure 2b**) [33]. Note, the increased coupling strength adversely affects the stability of the laser. As illustrated in **Figure 2c** light-versus-current characteristics start displaying 'kinks' while the coupling strength increases from $\eta =0.12$ to $0.3$.

**Figure 3** shows a top view of the fabricated hexagonal VCSEL and the calibrated 45 GHz small signal measurement test-setup. This top emitting VCSEL structure is grown by metal organic chemical vapor deposition (MOCVD) on an $n^+$ substrate. The epitaxial structure consists of 35 pairs of Si-doped bottom $Al_{0.16}Ga_{0.84}As/Al_{0.9}Ga_{0.1}As$ distributed Bragg reflector (DBR), while the cavity consists of three 70 Å, $In_{0.3}Ga_{0.7}As$-GaAs quantum wells (QWs) and a top 24-period DBR mirror. Inductively coupled plasma (ICP) etched the mesa into a semiconductor heterostructure, and its etch size was selected to be 2 µm larger than the active diameter (mesa diameter and oxidation time optimization). The aperture mesa diameters employed are 3.5 µm wide to ensure single transverse mode operation with a lower threshold current in the 3-mA pumping range. The inner cavity is an RF-modulated VCSEL, and the six identical outer cavities, adiabatically coupled to it provide parametrically selective slow light feedback. The end interface of each feedback cavity acts as a perfect mirror in the lateral direction, supporting the lateral optical light coupled back into the inner one [34].

However, the TCC-based VCSEL design has two fundamental challenges; (i) a typical limitation of the VCSEL with only a single TCC is the need for a strong slow-light coupling into the VCSEL cavity to enhance the modulation, and (ii) the TCC(s) add non-linearity impacting the modulation performance [18-21].

To mitigate or entirely avoid both possible limitations, we propose and show the use of a multitude of TCCs, rather than relying on a single cavity, but in such a way to induce direct slow-light feedback from each TCC. The small-signal frequency response ($S_{21}$) of the VCSEL was obtained by generating a low power modulating signal with a vector network analyzer (VNA). The output modulated intensity from the inner (laser) cavity with a fixed current is then collected via a single mode fiber (SMF), while all other cavities are operated below threshold current (2 mA). A high-speed photodetector (PD) collects the VCSEL's RF output and compares it to the original modulating source. The IM response showed enhanced CPR and multiple PPR before its 3-dB roll-off. The 3-dB roll-off of the MTCC-enhanced laser exceeds the conventional design by about five-fold compared with a conventional VCSEL fabricated on the same epi-wafer (**Figure 4a**). While the conventional VCSEL being driven at $I_{conventional}$ =7 mA, which is the maximum power before it saturates, our designed VCSEL is just operated at 8 mA for the inner cavity with surrounding cavities driven at constant current of 2 mA below threshold bias current. Even at this level, our 3-dB roll off is beyond the photo-detector limit (> 45 GHz). **Figure 4b** depicts measured L-I curve for our TCC VCSEL with an effective aperture size of 3.5×25 µm$^2$, and a conventional VCSEL fabricated on the same epitaxial wafer with an aperture size of 3×3 µm$^2$. Interestingly, with our hexagonal VCSEL driving more current even beyond the conventional limitation (> 7 mA), output power linearly increases to about 5.5 mW, which is almost triple of its conventional VCSEL. It is also, important to mention that the threshold current of our hexagonal VCSEL can be further reduced via optimization of the oxide layer structure close to conventional VCSEL. Thanks to the Vernier effect in the MTCC VCSEL, even with such a large oxide aperture, a single mode operation with SMSR of > 30 dB (7dB more than conventional VCSEL) [35] and SNR of > 45 dB are obtained **(Figure 4c)**. Note, that to obtain *η~0.45* (which potentially provide a 3-dB roll-off ~ 100 GHz), one can design oxide-free VCSELs recently reported by Deppe [36].

## 4. Conclusion

In conclusion, we propose a novel design of a 980 nm VCSEL adiabatically- and laterally coupled to six hexagonal feedback cavities. Succeeding this approach, we demonstrate a 5-fold higher 3-dB roll-off laser modulation bandwidth (>45 GHz limited by the experimental setup) compared to a non-coupled, conventional design.

This coupled hexagonal VCSEL paradigm shows single-mode operation with SMSR> 30 dB, which is 7 dB higher than conventional VCSELs fabricated on the same epi-wafer. Furthermore, with an SNR of > 45 dB, the peak output power of 5.5 mW is about triple as high compared to the conventional design. Further bandwidth enhancement for the VCSEL with MTCC is the need for a strong slow-light coupling into the modulating cavity. For instance, to obtain *η~0.45*, which potentially provides 3-dB roll-off ~ 100 GHz. This device concept is promising platform for the next generation of optical interconnects.

**Funding**
Air Force Office of Scientific Research (AFSOR) Small Business Innovation Research (SBIR) Program (FA9550-19-C-0003).

**Acknowledgements**
We thank Dr. Gernot Pomrenke and Dr. Chung-Yi Su for their fruitful discussions.

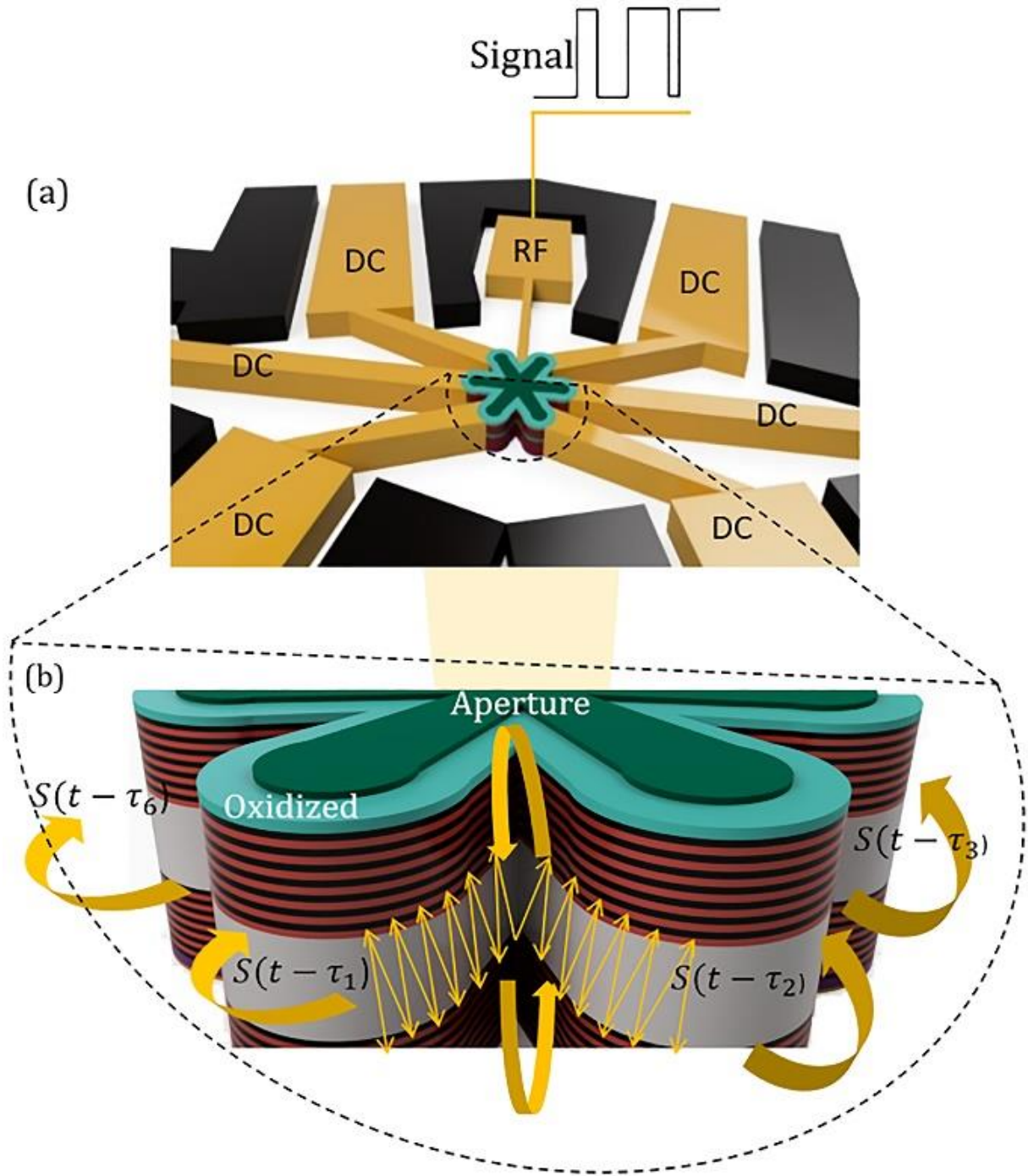

**Figure 1.** Schematic structure of our hexagonal transverse coupled cavity VCSEL (a) Top view, and (b) Cross section view. The six outer cavities are designed to have the same resonant wavelength ω at $I_{feedback} < I_{th}$. $I_{laser} = mI_{th}$. When $I_{laser}$ increases resonant wavelength red shift like its conventional VCSEL.

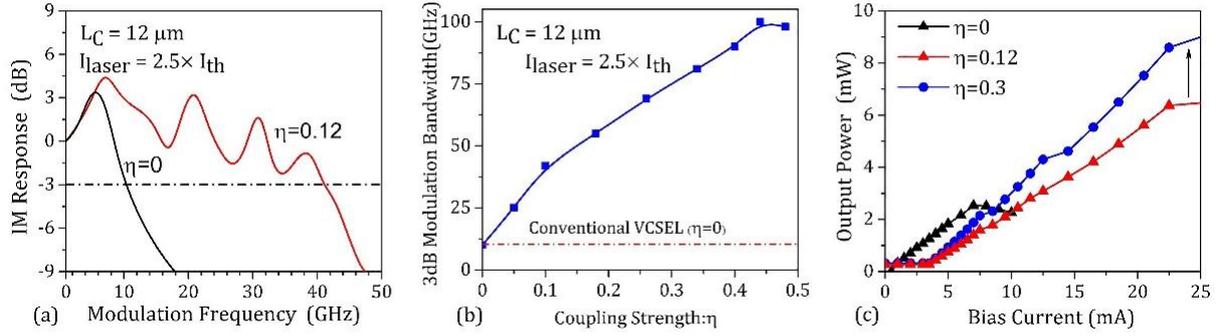

**Figure 2.** The design of hexagonal transverse coupled cavity VCSEL. (a) Small-signal IM response for VCSEL with six identical feedback cavities with length and coupling strength of $L_C$=12 µm and *η=0.12*, respectively. IM response with a robust feedback system (*η=0.12*) provides 3-dB roll-off more than 42 GHz (Equation 1 and Equation 2). Conventional VCSEL fabricated on the same epi-wafer with 3-dB roll off ~9 GHz is shown for comparison (*η=0*), (b) Shows the 3-dB modulation bandwidth versus coupling strength (*η*). The 3-dB roll-off can exceed 100 GHz via further increasing the coupling strength to η=0.45. Note, the increased of the coupling strength directly affects the stability of the laser, and (c) shows output power as a function of bias current of the inner cavity. The black line with a triangle symbol indicates the L-I curve of a conventional VCSEL with an aperture size of 3 × 3 µm². The Red and Blue curves are for MTCC VCSEL with six identical feedback cavities and $L_C$=12 µm. It is noted that linearity of the L-I curve deteriorates as coupling strength increases from η=0.12 to 0.3. Also, there is a trade-off between wall-plug efficiency and kinks in the L-I curve.
12

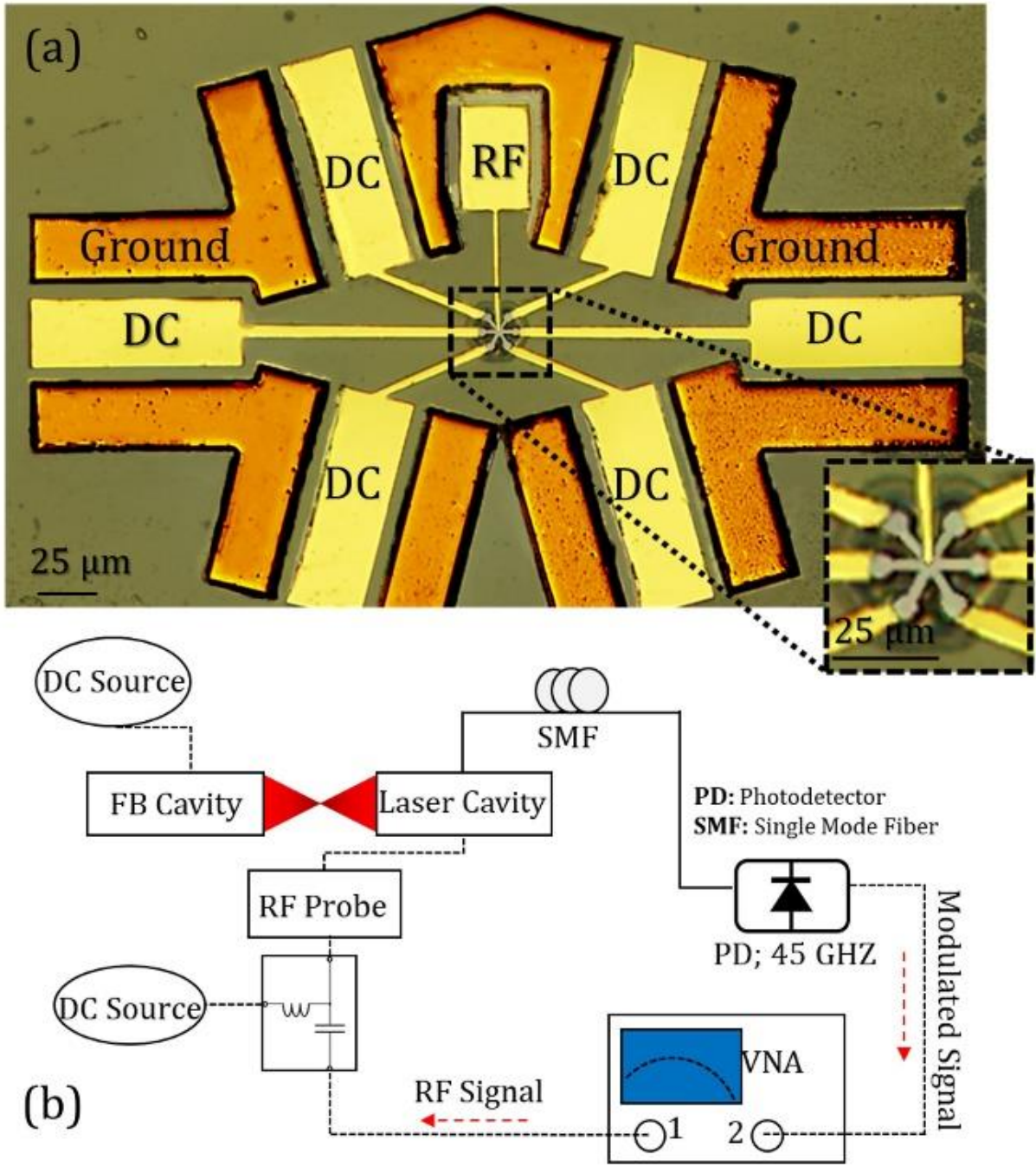

**Figure 3.** (a) The top-view false optical microscopic image of the fabricated MTCC VCSEL with aperture size of 3.5 × 3.5 µm$^2$ for inner cavity and six identical feedback, each with a length of 12 µm. With an aim to achieve high power, single-mode operation, the Vernier effect in lateral integration in a VCSEL can be utilized to form a larger aperture size and hence, more gain medium [25,26], and (b) our calibrated 45 GHz small signal measurement setup.

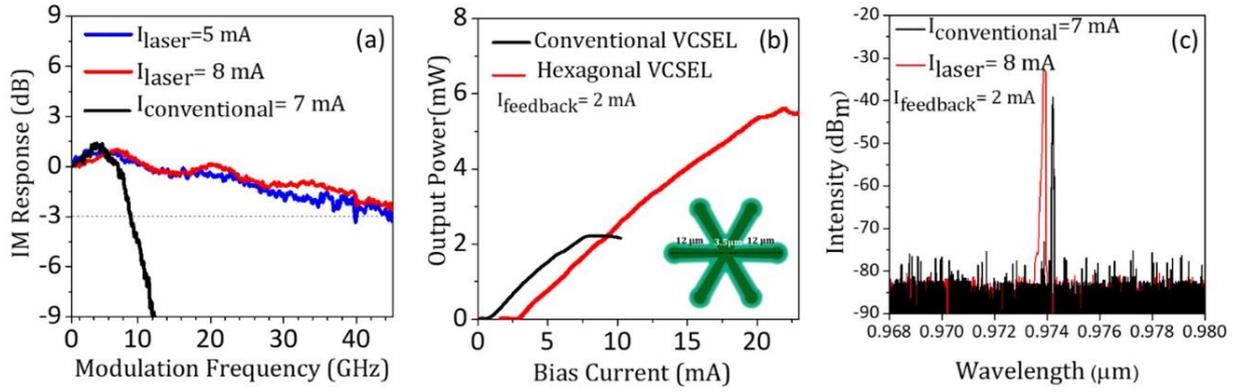

**Figure 4.** (a) IM response for VCSEL with six identical feedback cavities with the length of $L_C$=12 µm and effective aperture size of 3.5×25 µm$^2$. The IM response with a robust feedback system in the hexagonal VCSEL provides a 3-dB roll-off over 45 GHz based on Equation 1 and Equation 2. We project the coupling strength to be η=0.12. Conventional VCSEL fabricated on the same epi-wafer with 3-dB roll off ~9 GHz is shown for comparison, (b) shows out-put power for both conventional (black curve) and our hexagonal one (red curve) versus bias current. As seen here, our hexagonal VCSEL can be driven with large of currents (> 20 mA) and output power still linearly increases to ~ 5.5 mW, which is triple of its conventional VCSEL, and (c) a single mode operation for our hexagonal VCSEL with SMSR of > 30 dB and SNR of > 45 dB is obtained.